\documentclass[10pt]{iopart}
\usepackage{graphicx}
\usepackage{float}
\usepackage[section]{placeins}
\usepackage{amssymb}
\usepackage{appendix}
\usepackage{cite}
\usepackage{xcolor}

\begin{document}

\title[Superconducting Vortices]{Interaction Between Two Single Superconducting Vortices Inside A Superconducting Hollow Cylindrical domain}

\author{D. Garc\'ia Ovalle$^1$, E. Mu\~noz$^2$ and R. D. Benguria$^3$}

\address{Faculty of Physics, Pontificia Universidad Católica de Chile, Avda. Vicu\~na Mackenna 4860, Santiago, Chile}
\ead{\mailto{ddgarcia@uc.cl}$^1$, \mailto{munozt@fis.uc.cl}$^2$,\mailto{rbenguri@fis.uc.cl}$^3$}
\vspace{10pt}
\begin{indented}
\item[]December 2, 2019
\end{indented}

\begin{abstract}
Inspired by the seminal, ground-breaking work of Abrikosov in 1957, we developed a new approximation to the interaction between two widely separated superconducting vortices. In contrast with Abrikosov's, we take into account the finite size of the vortices and their internal magnetic profile. We consider the vortices to be embedded within a superconducting, infinitely long hollow cylinder,
in order to simplify the symmetry and boundary conditions for the mathematical analysis. We study this system in the context of a magnetic Ginzburg-Landau functional theory, by solving for the magnetic field profile inside each vortex, as well as in the superconducting region, subject to physical boundary conditions inspired by the classical analogue of two mutually inducting coils. Under isothermal conditions, the effective force between these vortices is given by the gradient of the Helmholtz free energy constructed from the Ginzburg-Landau functional. From our results, we explicitly show that, in agreement with well established theoretical arguments and experiments, the interaction between widely separated vortices is repulsive in this context,  and their equilibrium positions are constrained by the fluxoid's conservation. Moreover, we find that the equilibrium positions of the vortices' centers are stable due to the convexity of the Helmholtz free energy profile. Remarkably, the effect of the boundaries of the region over the effective interaction between the vortices is important in the chosen geometric configuration. 
\end{abstract}



\maketitle
\ioptwocol 
\section{Introduction}\label{I}

In conventional superconductivity,the phenomenological magnetic Ginzburg-Landau model reproduces the macroscopic behavior of superconducting samples near their critical temperature $T_c$ \cite{Ginzburg}. In particular, this model allows us to understand the physical behavior of vortices in these samples.  

Superconducting vortices are related to the flux quantization (or fluxoid quantization in non-bulk samples) phenomena predicted by F.~London $\&$ H.~London in 1950~\cite{London} and corroborated by Onsager~\cite{Onsager}, Bardeen~\cite{Bardeen} and Byers $\&$ Yang~\cite{Byers} in 1961. In the same year, experimental evidences about these objects were found by Deaver $\&$ Fairbank~\cite{Deaver} and Doll $\&$ N\"abauer\cite{Doll}. In the context of superconductivity, vortices can be described as regions where the fluxoid is quantitatively important, due to the low mean density of superconducting electrons inside the sample. These kind of quantum vortices are the main phenomena in Type II superconductivity, where the Helmholtz free energy is minimized by increasing the number of them. 

Abrikosov shows in his seminal work of 1957\cite{Abrikosov}, in the context of cylindrical symmetry, that in the extreme Type II case $\kappa=\lambda\xi^{-1}>>1$, where $\xi=\hbar(2m^*|\alpha|)^{-1/2}$ is the coherence length and $\lambda=(4\pi (q^*)^2\psi_{\infty}^2(m^*c^2)^{-1})^{-1/2}$ is the London penetration depth~\cite{Tinkham}, that the interaction between vortices can be explained within an approximation where they are considered as perturbations of the sample, neglecting their internal structure and boundaries. In Abrikosov's approach~\cite{Tinkham}, a small vortex centered at $\vec{x} = 0$ is described as a filament with negligible radius $\xi \rightarrow 0$ that, nevertheless, concentrates a finite fluxoid $\Phi_0=hc(q^*)^{-1}$ at its center. Therefore, the magnetic field $\vec{B}_A$ inside the sample is assumed to satisfy the London equation \cite{London}: 

\begin{equation}
    \nabla^2\vec{B}_A-\frac{\vec{B}_A}{\lambda^2}=-\frac{\Phi_0\delta_2(\vec{x})}{2\pi\lambda^2}\hat{k}, \label{1}
\end{equation}

\noindent 
where $\delta_2(\vec{x})$ is a two dimensional delta-function describing the concentration of the fluxoid at the center of the vortex. The explicit solution for Eq.~(\ref{1}) is: 

\begin{equation}
    \vec{B}_A=\frac{\Phi_0}{2\pi\lambda^2}K_0\left(\frac{r}{\lambda}\right)\hat{k},\label{2}
\end{equation}

\noindent 
with $K_0(x)$ the modified Bessel function of the second kind and zero order. For $\kappa>>1$, if $\vec{x}_1$ and $\vec{x}_2$ are the locations of the small vortices, the magnetic field at the position $\vec{x}$ in the system is given by the superposition of the magnetic fields generated by each of them:

\begin{equation}
    \vec{B}(\vec{x})=\left(\vec{B}_A(|\vec{x}-\vec{x}_1|)+\vec{B}_A(|\vec{x}-\vec{x}_2|)\right)\hat{k}.\label{3}
\end{equation}

In this approximation, the vortex energy per unit length is \cite{Tinkham}: 

\begin{equation}
    \epsilon=\frac{1}{8\pi}\int(|\vec{B}|^2+\lambda^2|\nabla\times \vec{B}|^2)dS,\label{4}
\end{equation}
and combining Eq.~(\ref{2}) and Eq.~(\ref{3}), the interaction energy per unit length between both vortices is

\begin{equation}
\epsilon_{12}=\frac{\Phi_0^2}{8\pi^2\lambda^2}K_0\left(\frac{|\vec{x}_1-\vec{x}_2|}{\lambda}\right).\label{5}
\end{equation}

The interaction between widely separated vortices must be repulsive in Type II superconductivity, since the contribution of the magnetic energy is bigger than the effects of the quantum currents~\cite{Kramer}. Theoretically, this behavior is also suggested by the Abelian Higgs model \cite{Rebbi,Speight} and the Boson method applied to the study of vortex lines~\cite{Leplae}. This fact is also observed experimentally~\cite{Hove,Sow} and numerically~\cite{Chaves,Mohamed}.

We suggest a new approximation to the interaction between two single superconducting vortices, inside a superconducting domain with the shape of an infinitely long hollow cylinder. We choose this geometry for two reasons: First, it represents the cross section of a long and thin superconducting coaxial cable, which is suitable for experimental applications. Second, but not less important, the cylindrical symmetry of the domain simplifies the calculations related to the boundary conditions, which uniquely define the magnetic field at each vortex and in the superconducting region and, as we show later, are essential in determining the effective force. 

We propose an ansatz for the order parameter, and we solve the magnetic field inside each vortex as well as inside the superconducting region, subjected to physical boundary conditions. The main feature of this approach is to recognize the contribution of the magnetic structure of each vortex and the superconducting region. In this sense, our model employs the electrodynamic analogue for the problem of two mutually inducting coils, where the magnetic flux inside the first coil is in part produced by the second coil, and viceversa. The magnetic field inside each vortex is determined by the boundary conditions related to the regularity of the magnetic vector potential, the continuity of the magnetic field inside and outside each vortex, and a self consistent solution for the magnetic field and the magnetic flux inside each vortex. We neglect the small physical effects of the vortices over the coaxial cylindrical boundaries, in order to preserve mathematical simplicity. Besides, each vortex is assumed to be submitted to the magnetic field imposed by the superconducting region and by the other vortex. 

Under isothermal conditions, the effective force between the small vortices is determined as the gradient of the Helmholtz free energy. Due to the complexity of the analytical expressions, a numerical evaluation of these results is shown in Fig.~2 -- Fig.~13, considering vortices with quantum currents circulating in the same direction, as well as in opposite directions.  

Our article is organized as follows: In Section 2, we present the context of the problem and we describe our strategy for its solution. In Section 3, we calculate self-consistently the magnetic vector potential and the magnetic field inside each vortex and within the superconducting region. In Section 4, we show the general form of the Helmholtz free energy profile  and the effective force acting on each vortex. In Section 5, due to the complexity of the expressions for the energy profile and the force, we develop a numerical evaluation of our analytical results, with plots that illustrate the physical behavior of the vortices.  


\section{The Interaction Problem}\label{II}

Let us consider a superconducting region with the shape of an infinitely long hollow cylinder, with internal and external radii $R_0 < R$,  respectively. We further assume that this sample contains two identical single vortices, with radius $\xi$ in the $\kappa>>1$ limit. An external magnetic field $\vec{H}_0$ is applied to the sample, with $H_{p} \leq H_0\leq H_{u}$. Here, $H_{p}$ and $H_{u}$ are the first and the upper critical magnetic fields, respectively, for type II superconductivity. These critical fields are straightforward to obtain in the $\kappa>>1$ limit~\cite{Annett,Bennemann,Svistunov,Tinkham}. 

\begin{figure}[h]
\centering
\includegraphics[width=8cm]{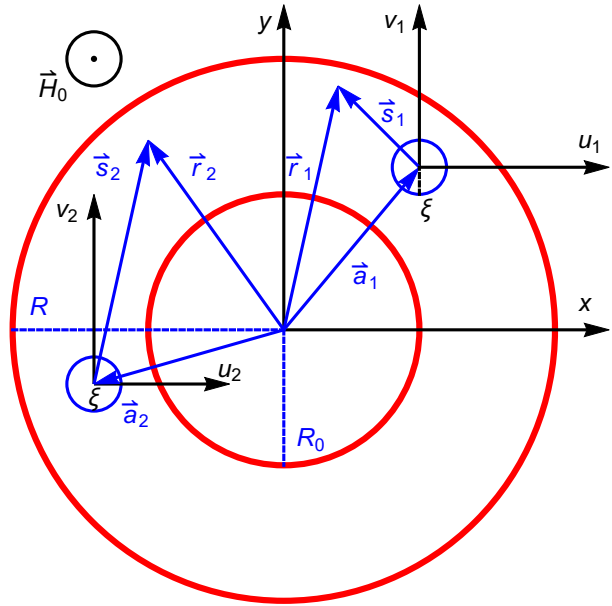}

\caption{Two vortices inside a superconducting, hollow cylindrical domain. The unit vectors $\hat{i}$, $\hat{j}$, $\hat{k}$, describe the usual basis in cartesian coordinates.}
\label{fig1}
\end{figure}

The local coordinate system for each vortex ($k=1,2$), as illustrated in Fig.~\ref{fig1}, is determined by the following vector relations: 
\begin{numparts}
\begin{eqnarray}
\vec{r}_k &=&\vec{s}_k+\vec{a}_k = r_k(\cos\theta_k\hat{i}+\sin\theta_k\hat{j}), \label{6a}
\end{eqnarray}
with
\begin{eqnarray}
\vec{s}_k&=&s_k(\cos\phi_k\hat{i}+\sin\phi_k\hat{j}),\nonumber\\
\vec{a}_k&=&a_k(\cos\alpha_k\hat{i}+\sin\alpha_k\hat{j}),
\label{6b}
\end{eqnarray}
and where we have defined the unitary vectors
\begin{eqnarray}
\hat{\theta}_k&=&-\sin\theta_k\hat{i}+\cos\theta_k\hat{j},\nonumber\\
\hat{\phi}_k&=&-\sin\phi_k\hat{i}+\cos\phi_k\hat{j},\nonumber\\
\hat{\alpha}_k&=&-\sin\alpha_k\hat{i}+\cos\alpha_k\hat{j}.\label{6c}
\end{eqnarray}
\end{numparts}

Considering the following definitions: 
\begin{eqnarray}
\Omega&:=&\{\vec{r}\in \mathbb{R}^2 \left.\right| R_0<|\vec{r}|<R \},\label{7}\\
\Omega_k&:=&\{\vec{s}_k\in \mathbb{R}^2 \left.\right| |\vec{s}_k|<\xi\},  \qquad k=1,2, \label{8}
\end{eqnarray}

\noindent 
the effective force acting on the vortex $\Omega_k$, under isothermal and reversible conditions, is given by

\begin{equation}
    \vec{f}_k=-\nabla_{\vec{a}_k} F\qquad k=1,2.\label{9}
\end{equation}

In equation (\ref{9}), $F$ is the Helmholtz free energy in the magnetic Ginzburg-Landau model, expressed in gaussian units  \cite{Tinkham}: 

\begin{equation}
F=\sum_{k=1}^2\int_{\Omega_k}\mathcal{F}d^2 s_k + \int_{\Omega\setminus(\Omega_1\cup\Omega_2)}\mathcal{F}d^2 r,
\label{10}
 \end{equation}
 
\noindent 
with the functional
\begin{eqnarray}
\mathcal{F}&=&\alpha|\psi_k|^2+\frac{\beta|\psi_k|^4}{2}+\frac{\left|\left(\frac{\hbar}{i}\nabla_k-\frac{q^*\vec{A}_k}{c}\right)\psi_k\right|^2}{2m^*}\nonumber\\
&+& \frac{|\vec{B}_k|^2}{8\pi} \qquad k=0,1,2, \label{11}
\end{eqnarray}
and the gradient in the coordinates defined by Eq.~(\ref{6b}),
\begin{eqnarray}
\nabla_k&=&\frac{\partial}{\partial s_k}\hat{s}_k+\frac{1}{s_k}\frac{\partial}{\partial \phi_k}\hat{\phi}_k \qquad k=1,2. \label{12}
\end{eqnarray}
For each vortex $\Omega_k$, for $k=1,2$, the order parameter $\psi_k$ and the magnetic vector potential $\vec{A}_k$ depend on the cylindrical coordinates $(s_k,\phi_k)$, while inside the hollow cylindrical region $\Omega\setminus(\Omega_1\cup\Omega_2)$, we denote these quantities with the $k=0$ index. Looking for a saddle-point of the energy functional, 
\begin{eqnarray}
\frac{\delta F}{\delta\psi_{k}^{*}} = 0,\qquad k = 0,1,2
\label{13}
\end{eqnarray}
we obtain the Ginzburg-Landau equations for the order parameters $\psi_k$ on each region~\cite{Tinkham}
\begin{eqnarray}
\left[\frac{\left(-i\hbar\nabla_k - \frac{q^*}{c}\vec{A}_k\right)^2}{2 m^*} + \alpha + \beta|\psi_k|^2\right]\psi_k = 0.
\label{14}
\end{eqnarray}
Similarly, a saddle point of the functional with respect to the vector potential components 
\begin{eqnarray}
\frac{\delta F}{\delta  \vec{A}_k}=0, \qquad k=0,1,2
\label{15}
\end{eqnarray}
leads to a generalization of Ampere's law~\cite{Tinkham}
\begin{eqnarray}
    \frac{c}{4\pi}\nabla_k\times \vec{B}_k&=&\frac{q^*\hbar(\psi_k^*\nabla_k\psi_k-\psi_k \nabla_k\psi_k^*)}{2m^*i}\nonumber\\
    &-&\frac{(q^*)^2|\psi_k|^2\vec{A}_k}{m^*c}\qquad k=0,1,2. \label{16}
\end{eqnarray}

Solutions for Eq.~(\ref{14}) and Eq.~(\ref{16}) are unique with physically appropriate boundary conditions. These conditions contain the information for the interaction between vortices, and involve the magnetic field and the corresponding magnetic flux in $\Omega\setminus(\Omega_1\cup\Omega_2)$ in a self-consistent way, as we shall later explain in detail. The magnetic field in the superconducting domain is obtained by neglecting the effects of the vortices at the boundary of the sample, within a mean field approximation to the problem. 


\section{Order Parameters and Magnetic Fields for the Sample}\label{III}

\subsection{Order Parameter and Magnetic Field for the Superconducting Region.}\label{IIIA}
The region $\Omega\setminus(\Omega_1\cup\Omega_2)$ is  superconducting. Therefore, we assume that this domain is in the Meissner state, and hence an ansatz for the order parameter $\psi_0$, considering one fluxoid quantum is \cite{Tinkham}:

\begin{equation}
    \psi_0=\psi_{\infty}\exp(i\theta), \qquad \psi_{\infty}=\sqrt{-\frac{\alpha}{\beta}}.\label{17}
\end{equation}

\noindent 

Using Eq.~(\ref{17}), the fundamental relation $\vec{B}_0=\nabla\times \vec{A}_0$ and Coulomb's gauge $\nabla\cdot\vec{A}_0=0$, Eq.~(\ref{16}) can be solved for $\vec{A}_0\in \Omega\setminus(\Omega_1\cup\Omega_2)$ within the geometry described in Fig.~\ref{fig1}. As shown in detail in \ref{B}, the general expressions for the magnetic vector potential $\vec{A}_0$ and the magnetic field $\vec{B}_0$, inside the superconducting domain, are given in terms of modified Bessel functions:

\begin{eqnarray}
    \vec{A}_0&=&\left(c_1I_1\left(\frac{r}{\lambda}\right)+c_2K_1\left(\frac{r}{\lambda}\right)+\frac{\Phi_0}{2\pi r}\right)\hat{\theta}, \label{18}
\end{eqnarray}

\begin{eqnarray}
    \vec{B}_0&=&\frac{1}{\lambda}\left(c_1I_0\left(\frac{r}{\lambda}\right)-c_2K_0\left(\frac{r}{\lambda}\right)\right)\hat{k}. \label{19}
\end{eqnarray}

\noindent 
Here, $c_1$ and $c_2$ are constants that depend on the boundary conditions (see \ref{B}). On the other hand, the magnetic field must be continuous at $r=R_0$ and $r=R$. If we take into account that the external magnetic field is constant outside $\Omega$, and assuming that the effects of each vortex at the coaxial cylindrical boundaries are sufficiently small to be neglected for $\kappa \gg 1$ ($\xi \ll \lambda < R - R_0$), then the boundary conditions are

\begin{eqnarray}
    B_0(R_0) = B(R) = H_0. \label{20}
\end{eqnarray}

Then, in terms of the auxiliary functions
\begin{eqnarray}
g_{\pm}(R_0,R)&=&I_0\left(\frac{R_0}{\lambda}\right)\pm I_0\left(\frac{R}{\lambda}\right),\\
h_{\pm}(R_0,R)&=&K_0\left(\frac{R_0}{\lambda}\right)\pm K_0\left(\frac{R}{\lambda}\right),\nonumber\\
\mathcal{G}(R_0,R)&=&\frac{(g_+h_--g_-h_+)}{2H_0\lambda}\nonumber\label{21}
\end{eqnarray}
the constants $c_1$ and $c_2$ are given by

\begin{equation}
    c_1=\frac{h_-(R_0,R)}{\mathcal{G}(R_0,R)}, \qquad c_2=\frac{g_-(R_0,R)}{\mathcal{G}(R_0,R)}.\label{22}
\end{equation}

\subsection{Order Parameter and Magnetic Field for Each Vortex.}\label{IIIB}

In terms of the cylindrical coordinates related to each vortex $(s_k,\phi_k)$, for $k=1,2$, we develop a self consistent solution for the magnetic field at each vortex, that determines their internal and external profile. In this sense, we assume that each vortex is subjected to a superposition of the magnetic field produced by the superconducting, hollow cylindrical region, and the external profile of the magnetic field generated by the other vortex. The mathematical expression for this statement will be presented in detail when we describe the continuity and boundary conditions for the magnetic field in Section~\ref{IIIC}.

\subsubsection{External Profile of the Magnetic Field for Each Vortex.}\label{IIIB1}

The magnetic field generated by each vortex in the region $s_k>\xi$, as a solution of Eq.~(\ref{16}), has the general form:

\begin{eqnarray}
    \vec{A}_{k,E}=\left(d_{k,E}I_1\left(\frac{s_k}{\lambda}\right)+e_{k,E}K_1\left(\frac{s_k}{\lambda}\right)+\frac{n_k\Phi_0}{2\pi s_k}\right)\hat{\phi}_k, 
    \label{23}
\end{eqnarray}

\begin{eqnarray}
    \vec{B}_{k,E}=\left(d_{k,E}I_0\left(\frac{s_k}{\lambda}\right)-e_{k,E}K_0\left(\frac{s_k}{\lambda}\right)\right)\frac{\hat{k}}{\lambda}. 
    \label{24}
\end{eqnarray}

Here, for $k=1,2$, $n_k$ is the number of fluxoids piercing each vortex. Besides, $d_{k,E}$ and $e_{k,E}$ are constants that depend on the boundary conditions (for explicit expressions, see Appendix D), as will be discussed in the next section.

\subsubsection{Internal Profile of the Magnetic Field at Each Vortex.}\label{IIIB2}

For $s_k<\xi$ and $k=1,2$, the order parameter that describes its internal structure can be approximated, in the $\kappa>>1$ limit, for a winding number $n_k$ \cite{Bennemann,Svistunov,Serfaty,Tinkham}  by
\begin{equation}
    \psi_{k,I}=\psi_{\infty}\left(\frac{s_k}{\xi}\right)^{|n_k|}\exp(in_k\phi_k) \qquad k=1,2.\label{25}
\end{equation}
This ansatz shows that the density of superconducting electrons is zero at the center of each vortex, $s_k=0$, and increases to $\psi_{\infty}$ at $s_k=\xi$. With Eq.~(\ref{25}) into Eq.~(\ref{16}) and $\epsilon=\kappa^{-1}$, we show that the magnetic vector potential inside each vortex $\vec{A}_{k,I}$ satisfies the equation 

\begin{eqnarray}
\frac{-\Phi_0n_ks_k^{2|n_k|+1}}{2\pi\epsilon^{2|n_k|}\lambda^{2|n_k|+2}}&=&s_k^2A_{k,I}''+s_kA_{k,I}'-A_{k,I}\nonumber\\
&-&\frac{A_{k,I}s_k^{2|n_k|+2}}{\lambda^2\xi^{2|n_k|}} \qquad k=1,2\label{26}
\end{eqnarray}

\noindent
or, in terms of $w_k=s_k\lambda^{-1}$, for $k=1,2$, one obtains:

\begin{eqnarray}
-\frac{\Phi_0n_kw_k^{2|n_k|+1}}{2\pi\lambda\epsilon^{2|n_k|}}&=&w_k^2A_{k,I}''+w_kA_{k,I}'\nonumber\\
&-&A_{k,I}-\frac{A_{k,I}w_k^{2|n_k|+2}}{\epsilon^{2|n_k|}}.\label{27}
\end{eqnarray}

Equation (\ref{27}) can be solved using perturbative techniques \cite{Holmes,Serfaty} (For more details about this solution, see \ref{C}). Then, a perturbative solution for the magnetic vector potential and the magnetic field in $\Omega_k$, for $k=1,2$, is given by 

\begin{eqnarray}
\vec{A}_{k,I}&=&\left(\frac{d_{k,I}s_k}{\lambda\epsilon^{\tau_k}}+\frac{e_{k,I}\epsilon^{\tau_k}\lambda}{s_k}\right)\hat{\phi}_k \nonumber\\
&-&\left(\frac{\Phi_0n_ks_k^{2|n_k|+1}}{8\pi\lambda^2 |n_k|(1+|n_k|)\xi^{2|n_k|}}\right)\hat{\phi}_k,\label{28}
\end{eqnarray}
\begin{eqnarray}
\vec{B}_{k,I}&=&\left(\frac{2d_{k,I}}{\lambda\epsilon^{\tau_k}}-\frac{\Phi_0n_ks_k^{2|n_k|}}{4\pi\lambda^2|n_k|\xi^{2|n_k|}}\right)\hat{k},\label{29}
\end{eqnarray}

\noindent 
where $d_{k,I}$ and $e_{k,I}$ are constants that depend on the boundary conditions (for explicit expressions, see Appendix D), as will be discussed in the next section.


\subsection{Boundary Conditions.}\label{IIIC}

\subsubsection{Regularity of the Magnetic Vector Potential for Each Vortex.}\label{IIIC1}

We must discard divergent contributions at $s_k=0$ in Eq.~(\ref{28}). Therefore, we have 

\begin{equation}
    e_{k,I}=0 \qquad k=1,2. \label{30}
\end{equation}

\subsubsection{Continuity of the Magnetic Field.}\label{IIIC3}

The magnetic field at the boundary of each vortex $\partial \Omega_k$, for $k = 1,2$, must be continuous. Furthermore, by self-consistency, its value is given by the superposition of the magnetic field generated by the superconducting domain and the magnetic field produced by the other vortex,
\begin{eqnarray}
\lim_{\epsilon\rightarrow 0}\left.B_{1}\right|_{\partial \Omega_1^{-}} &=& \lim_{\epsilon\rightarrow 0}\left.B_{1}\right|_{\partial \Omega_1^{+}} = \left.B_{2}\right|_{\partial \Omega_1} + \left.B_{0}\right|_{\partial \Omega_1},\nonumber\\
\lim_{\epsilon\rightarrow 0}\left.B_{2}\right|_{\partial \Omega_2^{-}} &=& \lim_{\epsilon\rightarrow 0}\left.B_{2}\right|_{\partial \Omega_2^{+}} =  \left.B_{1}\right|_{\partial \Omega_2} + \left.B_{0}\right|_{\partial \Omega_2}.
\label{31}
\end{eqnarray}
Here, we defined $\partial\Omega_{k}^{-} = \mathcal{B}(\partial \Omega_k ,\epsilon)\cap\Omega_k$ and $\partial\Omega_{k}^{+} = \mathcal{B}(\partial \Omega_k ,\epsilon)\cap \Omega_{k}^c$, respectively,
with $\mathcal{B}(\partial \Omega_k ,\epsilon) =\left\{ \cup\,\mathcal{B}(\vec{\xi}_k,\epsilon),\,\,\vec{\xi}_k\in\partial\Omega_k \right\}$ the set of all possible balls of infinitesimal radius $\epsilon$, centered at any point at the boundary $\vec{\xi}_k\in\partial\Omega_k$.

From the system of coordinates displayed in Fig.~\ref{fig1}, the magnetic field due to the superconducting region at the boundary of each vortex can be expressed by
\begin{equation}
\left.B_0\right|_{\partial\Omega_k}=B_0\left(\frac{|\vec{a}_k+\vec{\xi}_k|}{\lambda}\right) \simeq B_0\left(\frac{|\vec{a}_k|}{\lambda}\right),\label{32}\\
\end{equation}
where $\vec{\xi}_k = \left.\vec{s}_k\right|_{s_k = \xi}$, following the definition in Eq.~(\ref{6b}).
Here, we have considered that in the $\kappa \gg 1$ limit, $\xi\lambda^{-1} \ll 1$, and hence $|\vec{a}_k+\vec{\xi}_k|\lambda^{-1}=(a_k^{2}+2 a_k\xi\cos\phi_k+\xi^2)^{1/2}\lambda^{-1}\sim |\vec{a}_k|\lambda^{-1}$. The same considerations imply that (for $k,k' = 1,2$)
\begin{eqnarray}
\left.B_{k}\right|_{\partial\Omega_{k' \ne k}}&  \simeq B_{k,E}\left(\frac{|\vec{a}_1-\vec{a}_2|}{\lambda}\right)\label{33}. 
\end{eqnarray}

Therefore, the continuity conditions stated in Eq.~(\ref{31}) can be reduced to the system of equations
\begin{eqnarray}
    B_{1,I}(s_1=\xi) &=& B_{1,E}(s_1=\xi) \nonumber\\
    &\simeq& B_{2,E}(|\vec{a}_1-\vec{a}_2|)
    + B_0(r=a_1), \nonumber\\
    B_{2,I}(s_2=\xi)&=& B_{2,E}(s_2=\xi)\nonumber\\
    &\simeq& B_{1,E}(|\vec{a}_1-\vec{a}_2|)
    +B_0(r=a_2).\label{34}
\end{eqnarray}

\subsubsection{Self Consistent Magnetic Flux.}\label{IIIC4}

The self-consistent continuity conditions for the magnetic field
stated in Eq.~(\ref{31}), whose approximate expression for $\kappa \gg 1$ is given by Eq.~(\ref{34}), imply similar considerations for the vector potential at the boundary of each vortex. It is convenient to express those conditions in terms of the circulation of the vector potential along the boundary of each vortex
\begin{eqnarray}
    \oint_{\partial\Omega_1}\vec{A}_{1,I}\cdot \vec{d}l&=&\oint_{\partial\Omega_1}(\vec{A}_0+\vec{A}_{2,E})\cdot \vec{d}l,\nonumber\\
    \oint_{\partial\Omega_2}\vec{A}_{2,I}\cdot \vec{d}l&=&\oint_{\partial\Omega_2}(\vec{A}_0+\vec{A}_{1,E})\cdot \vec{d}l.\label{35}
\end{eqnarray}
By Stokes' theorem, these equations state that the magnetic flux piercing the surface of each vortex is given by the superposition of the flux due to the field of the superconducting region, and the flux produced by the other vortex, in clear analogy with the classical model of two conducting, mutually inducting coils.

For $\kappa>>1$ and $\xi<<|\vec{a}_k|$, by similar considerations as those leading to Eq.~(\ref{34}), the boundary conditions in Eq.~(\ref{35}) can be written as the system of equations (For more details, see \ref{D}):
\begin{eqnarray}
\frac{2A_{1,I}(s_1=\xi)}{\xi}&\simeq&\frac{A_{0}(r=a_1)}{a_1}\nonumber\\
&+&\frac{A_{2,E}(s_2=|\vec{a}_1-\vec{a_2}|)}{|\vec{a}_1-\vec{a_2}|},\nonumber\\
\frac{2A_{2,I}(s_2=\xi)}{\xi}&\simeq&\frac{A_{0}(r=a_2)}{a_2}\nonumber\\
&+&\frac{A_{1,E}(s_1=|\vec{a}_1-\vec{a_2}|)}{|\vec{a}_1-\vec{a_2}|}. \label{36} 
\end{eqnarray}
The boundary conditions established in Eq.~(\ref{34}) and Eq.~(\ref{36}) allow us to determine all the constants leading to the complete solutions for the magnetic vector potential and the magnetic field. Due to the algebraic complexity of the equations, an application with the implementation of the boundary conditions for this model is shown in Section \ref{V}. Explicit analytical expressions for the constants are presented in \ref{Cnt}.


\section{General Form of the Helmholtz Free Energy and the Effective Force on Each Vortex}\label{IV}

With the order parameters, magnetic vector potentials and magnetic fields determined before, the Helmholtz free energy for the model can be expressed using Eq.~(\ref{10}) and Eq.~(\ref{11}) as follows:
\begin{eqnarray}
 F&\simeq&\sum_{k=1}^2\frac{d_{k,I}^2\epsilon^2}{2\epsilon^{2\tau_k}}\left(1+\frac{\epsilon^2}{4(2+|n_k|)}\right)-\frac{d_{k,I}\Phi_0n_k\epsilon^2}{8\pi\lambda(1+|n_k|)\epsilon^{\tau_k}}\nonumber\\
&-&\frac{d_{k,I}\Phi_0\epsilon^2 n_k}{8\pi\lambda(1+|n_k|)\epsilon^{\tau_k}}\left(\frac{1}{|n_k|}+\frac{\epsilon^2}{8|n_k|(1+|n_k|)}\right)\nonumber\\
&=& F_B + F_V, \label{37}
\end{eqnarray}
where the first term $F_B$ does not depend on the sign of the winding numbers $n_k$, while the second term $F_V$ does depend on it.
(See the computations of the relevant terms in \ref{E}). Using Eq.~(\ref{9}), with $\gamma=\cos(\alpha_1-\alpha_2)$, the effective force on the vortex $k$, for $k=1,2$, is given by 

\begin{eqnarray}
\vec{f}_{k}
&=&-\left(\frac{\partial  F}{\partial a_k}\hat{a}_k+\frac{(-1)^{k}\sqrt{1-\gamma^2}}{a_k}\frac{\partial  F}{\partial \gamma}\hat{\alpha}_k\right)\nonumber\\
&=& \vec{f}_{Bk}+\vec{f}_{Vk}.\label{38}
\end{eqnarray}
where we defined $\vec{f}_{Bk} = -\nabla_{\vec{a}_k} F_B$
and $\vec{f}_{Vk} = -\nabla_{\vec{a}_k} F_V$, respectively.

If we analyze the radial component of the effective force on each vortex, defined as
\begin{eqnarray}
f_{Bk}^{R} &=& \hat{a}_k\cdot\vec{f}_{Bk},\nonumber\\
f_{Vk}^{R} &=& \hat{a}_k\cdot\vec{f}_{Vk},
\label{38b}
\end{eqnarray}
we notice that $f_{V1}^{R} = f_{V2}^{R}$, for $n_1 = n_2$,
while $f_{V1}^{R} = - f_{V2}^{R}$ for $n_1 = - n_2$, thus yielding an effective attractive interaction for opposite winding numbers, and an effective repulsive interaction for identical winding numbers, respectively. However, since the total effective force is not only determined by this contribution, but also from the $f_{Bk}^{R}$ interaction defined in Eq.~(\ref{38}), that reflects the effects of the boundaries on each vortex, we can have a more complex scenario as discussed in the next section.

\section{Numerical Evaluation of the Results}\label{V}

\subsection{Previous Considerations.}\label{VA}

\subsubsection{Surface Energy.}

We remark that, in the limit $\kappa \gg 1$, the surface energy can be estimated at $H=H_C$, where $H_C$ is the thermodynamic critical field. Following the analysis shown in \cite{Fetter}, we can deduce that the surface energy of the system $\sigma_{ns}$ is approximately: 

\begin{eqnarray}
    \sigma_{ns}
    &=&\frac{H_C^2}{8\pi}\int_{\Omega}\left(\left\{1-\frac{B}{H_C}\right\}^2-\frac{|\psi|^4}{\psi_{\infty}^4}\right)d^2x\nonumber\\
    &\simeq&\frac{H_C^2\xi^2(1-\kappa)}{2}\left(2+\frac{(R+R_0)}{\xi}\right).\label{39}
\end{eqnarray}

As we can see from Eq.~(\ref{39}), $\sigma_{ns}\ll 0$. Therefore,
it is energetically favourable for the system to maximize its interfacial surface,
and hence to avoid for the vortices to attract each other and eventually coalesce. Hence, the thermodynamic analysis of the problem is consistent with an effective repulsive force between the vortices, as will be shown and discussed in the examples in Section~\ref{VD}.

Besides, the previous integral and the explicit forms of the magnetic fields and the order parameters show that the magnetic terms are the most important contribution to the surface energy (For more details, see \ref{F}). 

\subsubsection{Experimental Considerations.}
In Type II superconductivity, suitable values for the critical magnetic fields are given by $H_{p}=10^2\,G$ and $H_{u}=10^5G$\cite{Rohlf}, therefore $H_C=10^3\,G$. Besides, the fluxoid is given by $ \Phi_0=2.0679\times 10^{-7}G\,cm^2$ \cite{Tinkham,Ashcroft}. Finally, using the estimations for the critical magnetic fields mentioned before, we obtain that $\lambda=10^{-4}\,cm$ and $\xi=10^{-6}cm$, respectively. Therefore, for these parameters we estimate $ \kappa\simeq 100$. 

Concerning the typical sizes of the coaxial region, we notice that  in order to reproduce the effect of the London penetration depth, the internal and the external radii of the sample must satisfy $R-R_0> 2\lambda$. 
In addition, since we are exploring the strong influence of the magnetic profile in the superconducting region on the effective interaction between the vortices, we cannot impose a big difference between the radii of the coaxial cylinders. For all the previous reasons, we illustrate the model in the case $R_0=4\lambda$ and $R=8\lambda$. We represent the plots in terms of the dimensionless parameters: 

\begin{eqnarray}
    R'&=&\frac{R}{\lambda}, \qquad F'=\frac{10^3 \pi\,F}{AH_C^{2}},\qquad  a_k' = \frac{a_k}{\lambda} \nonumber\\
   f_k'&=&\frac{10^3 \pi\,f_k}{AH_C^{2}}, \qquad f_{Vk}'=\frac{10^4 \pi\,f_{Vk}^{R}}{AH_C^{2}}, \qquad k=1,2,\label{40} 
\end{eqnarray}
where $A = \pi \xi^2$ is the area of each vortex. 

\subsection{Superconducting current.}\label{current}
In order to understand the effective force over each vortex, it is instructive to first analyze the radial pattern of the current in the superconducting region $\Omega$. Here, we can identify two contributions to the total current: 

\begin{equation}
    \vec{J}_0=(J_s-J_d)\hat{\phi},\label{current_1}
\end{equation}
\noindent
where $J_s$ is the superconducting current and $J_d$ is the diamagnetic current. Firstly, for $J_s$ and using the order parameter $\psi_0 = \psi_\infty e^{i\theta}$:

\begin{eqnarray}
J_s&=&\frac{q^{*}}{2m^{*}i}\hat{\phi}\cdot\left(\psi_0^{*}\nabla\psi_0-\psi_0\nabla\psi_0^{*}\right),\nonumber\\
&=&\frac{q^{*}\hbar\psi_{\infty}^2}{m^{*}r}.\label{current_2}
\end{eqnarray}

For the diamagnetic current $J_d$, we have:

\begin{eqnarray}
    J_d&=&\frac{(q^{*})^2\psi_\infty^2A_0(r)}{m^{*}c}\nonumber\\
    &=&\frac{(q^{*})^2\psi_\infty^2}{m^{*}c}\left(c_1I_1\left(\frac{r}{\lambda}\right)+c_2K_1\left(\frac{r}{\lambda}\right)+\frac{\Phi_0}{2\pi r}\right).\label{current_3}
\end{eqnarray}

In terms of the dimensionless variables defined in Eq.~(\ref{40}), the total current is reduced to the expression

\begin{eqnarray}
    J_0'&=&\frac{4\pi\lambda J_0}{cH_C},\nonumber\\
    &=&-\left(c_1I_1(r')+c_2K_1(r')\right).\label{current_5}
\end{eqnarray}

\begin{figure}[h]
\includegraphics[width=1\columnwidth]{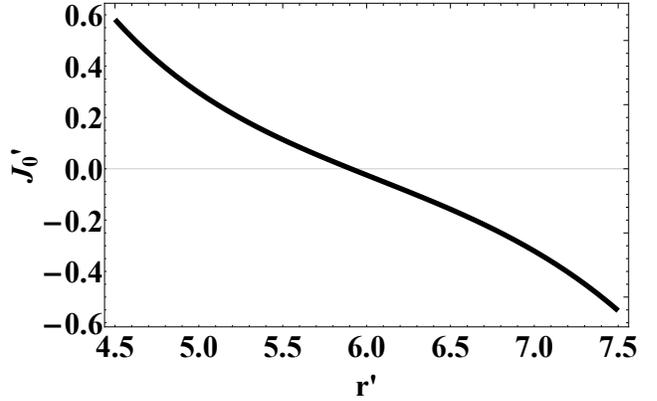}

\caption{The superconducting current in the hollow cylindrical region $\Omega$, as a function of the dimensionless radial distance $r'=r/\lambda$ from the center of the coaxial cylindrical boundaries.}\label{figure_current}
\end{figure}

In Fig.~\ref{figure_current}, we represent the total (dimensionless) current $J_0'$, as a function of the dimensionless radial distance $r'$, for a coaxial cylindrical sample of radii $R_0=4\lambda$ and $R=8\lambda$, respectively. As clearly seen in Fig.~\ref{figure_current}, the total current reverses its direction near $r' \sim 6.0$. This effect can be understood from a semiclassical picture after Amp\`ere's Law (and the corresponding right-hand rule), since the magnetic fields at the inner core and at the outer region have the same direction and magnitude, thus imposing a competition effect over the direction of the total current $J_0'$. This change of direction, as we shall discuss later, imposes a corresponding sign inversion on the dominant component of the radial effective force acting over the vortices.

\subsection{Helmholtz Free Energy Profile.}\label{VB}

\begin{figure}[h]
\includegraphics[width=1\columnwidth]{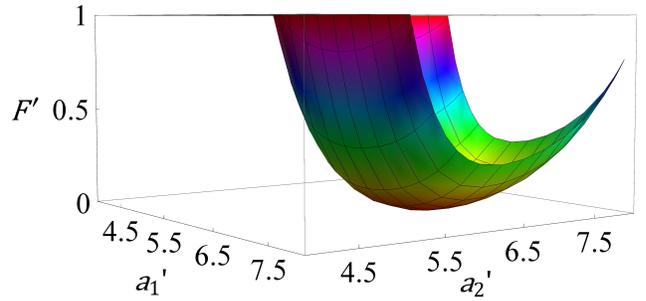}

\caption{Helmholtz free energy profile, in terms of the distance of the center of each vortex to the center of the coaxial cylinders. Here, $n_1=n_2=1$, $H_0=H_C$ and $\gamma=-1$.}\label{figure2}
\end{figure}

In Fig.~\ref{figure2}, the Helmholtz free energy is represented as a function of the (dimensionless) distance from the center of the coaxial cylinders to the center of each vortex, $a'_k$ for $k=1,2$. The relative angle is $\alpha_1 - \alpha_2 = \pi$, which implies $\gamma = \cos(\alpha_1 - \alpha_2) = -1$. Clearly, the functional is convex in terms of these variables, with a global minimum inside the cylindrical coaxial sample, that therefore represents the equilibrium position of the center of each vortex. In this example, the winding numbers of the two vortices are identical $n_1 = n_2 = 1$.
 
\begin{figure}[h]
\includegraphics[width=1\columnwidth]{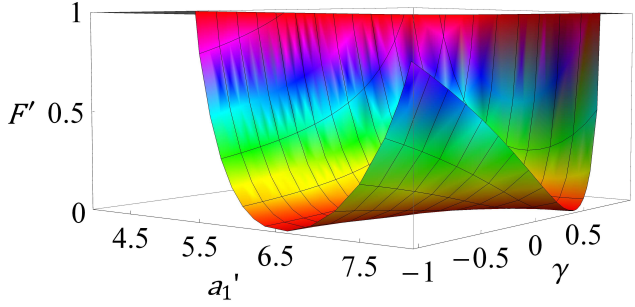}

\caption{Helmholtz free energy profile, in terms of the relative angle $\gamma = \cos(\alpha_1 - \alpha_2)$, and the distance to the center of the coaxial cylinders $a'_1$, where the symmetrical condition $a'_1=a'_2$ was chosen. Here, $n_1=n_2=1$ and  $H_0=H_C$.}\label{figure3}
\end{figure}

In Fig.~\ref{figure3}, the Helmholtz free energy is represented as a function of the relative angle $\gamma = \cos(\alpha_1 - \alpha_2)$, and the distance to the center of the coaxial cylinders $a'_1$, where the symmetrical condition $a'_1=a'_2$ was chosen. In this example, the winding numbers for each vortex are set identical $n_1 = n_2 = 1$. The free energy profile shows
a minimum at $\gamma = -1$, i.e. at $\alpha_1 - \alpha_2 = \pi$ where the
centers of the vortices are maximally separated, suggesting a repulsive interaction. We shall discuss this point in more detail in Section~\ref{VD},
after expressing the effective force. A similar behavior is observed when the winding numbers of the vortices are opposite, i.e. $n_1 = - n_2 = 1$.

\subsection{Radial Component of the Force on each Vortex.}\label{VC}\bigskip

\begin{figure}[h!]
\begin{centering}
\includegraphics[width=0.8\columnwidth]{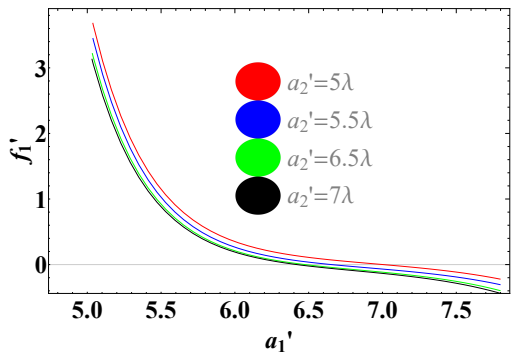}
\caption{Radial component of the force on the first vortex, as a function of its distance to the center of the coaxial cylinders, for different fixed positions of the second vortex. Here, $n_1=n_2=1$, $H_0=H_C$ and $\gamma=-1$. An analogue situation is obtained if the roles of the plot are exchanged.}\label{figure4}
\end{centering}
\end{figure}

From the information in Fig.~\ref{figure4}, the interaction between vortices with the same winding numbers and the boundary of the sample is repulsive. Besides, the fixed position of the second vortex displaces the effective radial force on the first vortex. This behavior is the same in the case of two vortices with opposite winding numbers, as it can be seen in the following plots: 

\begin{figure}[h]
\begin{centering}
\includegraphics[width=0.8\columnwidth]{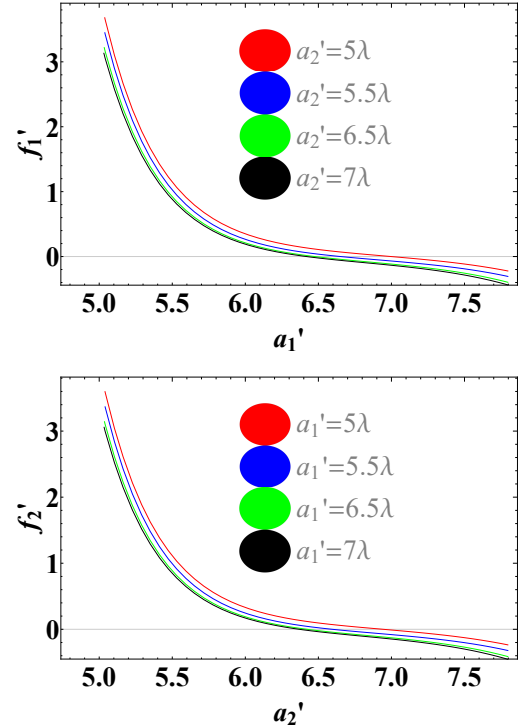}
\caption{Radial component of the force on the first vortex and the second vortex as a function of their distance to the center of the coaxial cylinders, for different fixed positions of the remaining vortex. Here, $n_1=1$, $n_2=-1$, $H_0=H_C$ and $\gamma=-1$.}\label{figure5}
\end{centering}
\end{figure}

\begin{figure}[h]
\includegraphics[width=1\columnwidth]{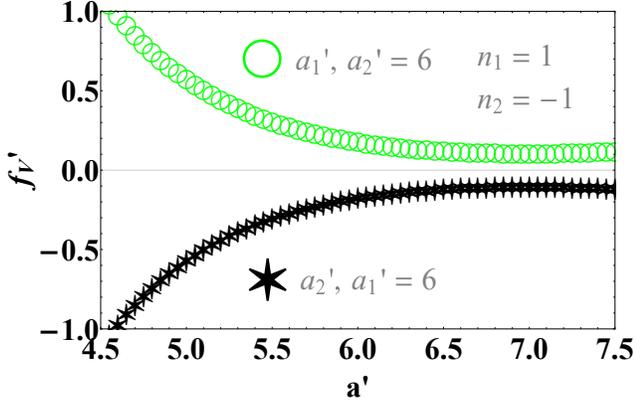}
\caption{Radial component $f_{Vk}^{R}$ of the force (as defined in Eq.~(\ref{38b}) and Eq.~(\ref{40})), in terms of the distance of the center of each vortex to the center of the coaxial region. Here, $n_1=n_2 = -1$, $H_0=H_C$ and $\gamma=-1$.}\label{figure_fV_neg}
\end{figure}

\begin{figure}[h]
\includegraphics[width=1\columnwidth]{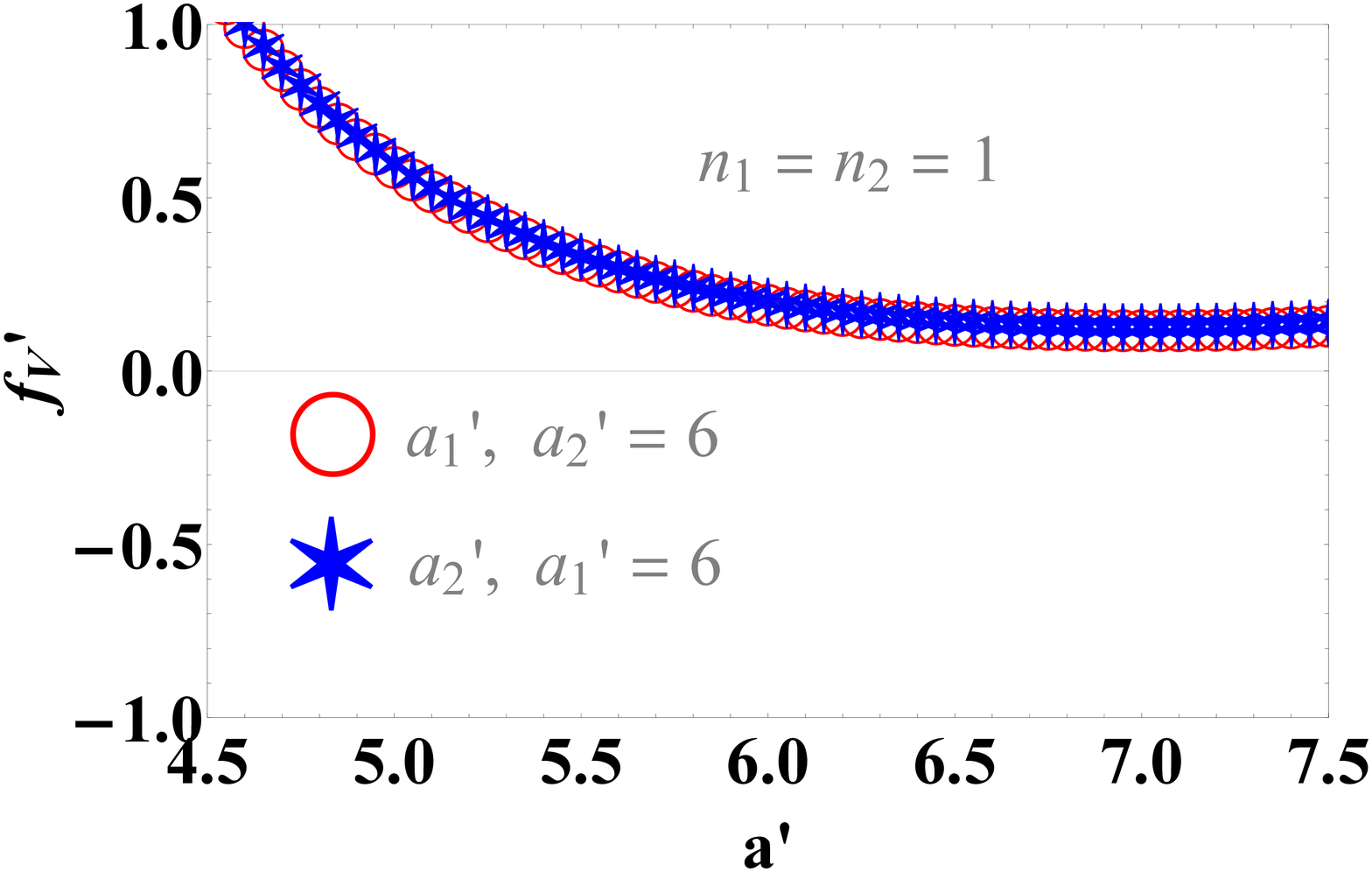}
\caption{Radial component $f_{Vk}^{R}$ of the force (as defined in Eq.~(\ref{38b}) and Eq.~(\ref{40})), in terms of the distance of the center of each vortex to the center of the coaxial cylindrical region. Here, $n_1=1$, $n_2 = 1$, $H_0=H_C$ and $\gamma=-1$.}\label{figure_fV_pos}
\end{figure}

\begin{figure}[h]
\includegraphics[width=1\columnwidth]{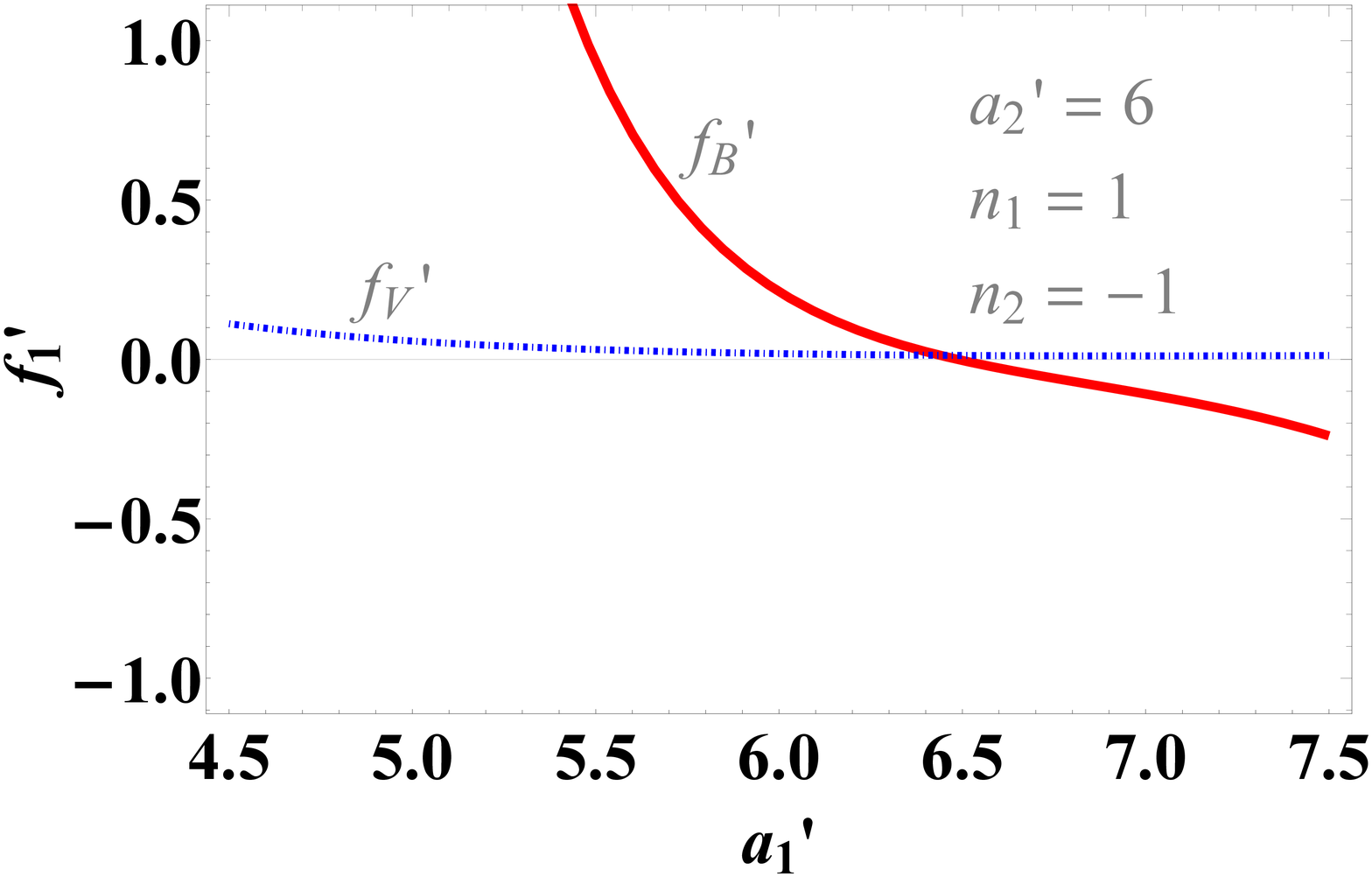}
\caption{Radial component of the force (as defined in Eq.~(\ref{38b}) and Eq.~(\ref{40})) acting on the first vortex. The plot shows the separate contribution of $f_{V1}^R$ and $f_{B1}^{R}$, respectively, as a function of the distance of the center of the vortex to the center of the coaxial cylindrical region. Here, $n_1=1$, $n_2 = -1$, $H_0=H_C$ and $\gamma=-1$.}\label{figure_fVB_1}
\end{figure}

\begin{figure}[h]
\includegraphics[width=1\columnwidth]{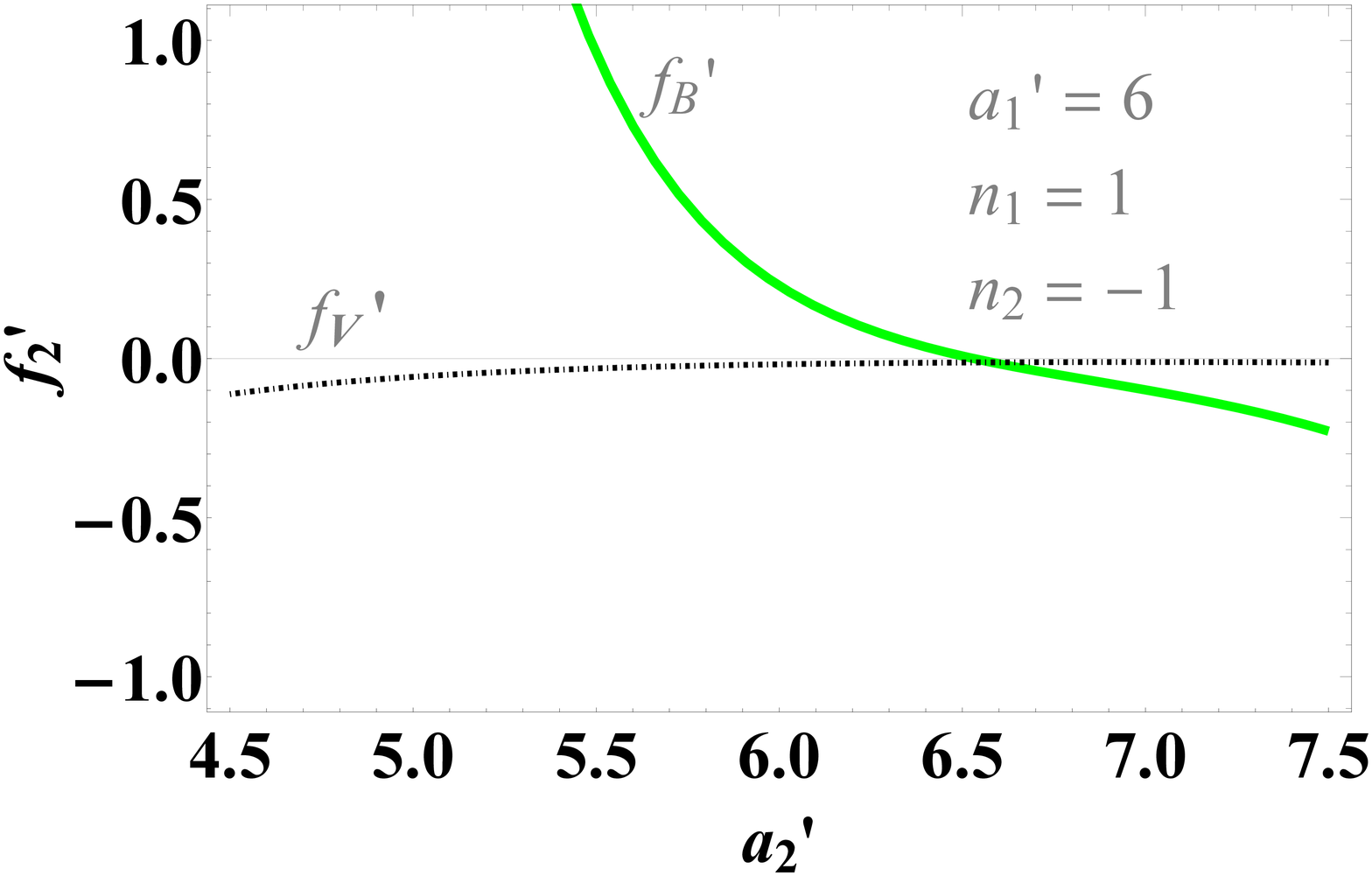}
\caption{Radial component of the force (as defined in Eq.~(\ref{38b}) and Eq.~(\ref{40})) acting on the second vortex. The plot shows the separate contribution of $f_{V2}^R$ and $f_{B2}^{R}$, respectively, as a function of the distance of the center of the vortex to the center of the coaxial region. Here, $n_1=1$, $n_2 = -1$, $H_0=H_C$ and $\gamma=-1$.}\label{figure_fVB_2}
\end{figure}

In order to understand this effect, it is better to analyze separately the two components of the radial force  defined in Eq.~(\ref{38b}) $f_{Vk}^R$ and $f_{Bk}^R$, respectively. As already discussed in section \ref{IV}, the component $f_{Vk}^{R}$ depends on the sign of the winding number $n_k$,
and thus reverts its relative sign for the case $n_1 = -n_2 = 1$ (see Fig.~\ref{figure_fV_neg}) as compared to the case $n_1 = n_2 = 1$ (see Fig. ~\ref{figure_fV_pos}). This sole contribution on itself would determine, as later discussed in section~\ref{VD} an attractive (repulsive) effective force between vortices with opposite (identical) winding numbers, respectively. However, the other contribution to the radial effective force $f_{Bk}^{R}$
does not depend on the sign of the winding numbers, since its value mainly represents the effect of the external field $H_0 = H_c$ imposed by the outer, normal regions, upon the superconducting region and the vortices themselves. The magnitude of the contribution $f_{Bk}^{R}$ of the net force over each vortex is displayed in Fig.~\ref{figure_fVB_1}
and Fig.~\ref{figure_fVB_2}, respectively. As clearly seen in these figures, for the parameter regime chosen where the boundaries of the sample are not too far, we have $|f_{Bk}^R| \gg |f_{Vk}^R|$, and hence the overall effective force over the vortices has the same direction for identical as well as opposite winding numbers, as seen in Fig.~\ref{figure4}
and Fig.~\ref{figure5}. Interestingly, an inversion of the direction (sign) of the dominant $f_{Bk}^{R}$ component occurs near $a'_k \sim 6.5$. This effect is correlative with the behavior of the current $J_0$, that reverses its direction close to this same distance.

\subsection{Interaction Between Vortices.}\label{VD}

From equation (\ref{38}), we calculate the interaction between vortices with the relative tangential component of the force, for $a=a_1=a_2$: 

\begin{eqnarray}
    \vec{f}_{12}&=&\frac{\sqrt{1-\gamma^2}}{a}\left(\frac{\partial F}{\partial \gamma}\right)(\hat{\alpha}_1+\hat{\alpha}_2).
    \label{41}
\end{eqnarray}
Now, with the change of variables:
\begin{eqnarray}
\chi_{\pm}&=&\cos\frac{\alpha_1\pm\alpha_2}{2},\label{42}
\end{eqnarray}

\noindent 
Eq.~(\ref{41}) can be written in the form 

\begin{eqnarray}
    \vec{f}_{12}&=&\frac{1}{2a}\left(\frac{\partial  F}{\partial \chi_-}\right)\sqrt{1-(2\chi_-^2-1)^2}\nonumber\\
    &&\left(-\sqrt{1-\chi_{+}^2}\hat{i}+\chi_{+}\hat{j}\right).\label{43} 
\end{eqnarray}

\begin{figure}[h]
\includegraphics[width=1\columnwidth]{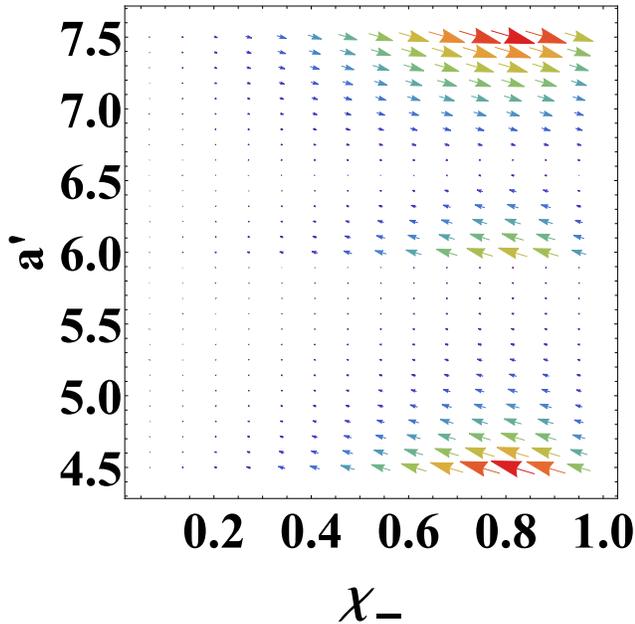}

\caption{Vector plot profile of the interaction between vortices, for $H_0=H_C$ and $n_1=n_2=1$ in terms of $a$ and $\chi_-$. Here, we fix that $\chi_+=(2)^{-1/2}$. The situation for vortices with opposite winding numbers are equivalent. Consider that the vortex's proximity to the outer boundary increases the influence of the sample, which explains the change in the sign of the force as the vortices go near $R$.}\label{figure6}
\end{figure}

The behavior of the force between vortices is illustrated in the vector field plot displayed in Fig.~\ref{figure6}, for $R/\lambda = 8.0$
and $R_0/\lambda = 4.0$. 

For $4.0 < a' < 6.5$, corresponding approximately to the condition $a < R_0 + (R-R_0)/2$, the influence of the outer boundary is weak as compared with the interaction between vortices, and hence Fig.~\ref{figure6} shows that $\chi_-=0$ ($\alpha_1 - \alpha_2 = \pi$) is an attractor for this situation, where the relative angle between vortices is maximum. Therefore, our model predicts a repulsive interaction between vortices in this limit. The interaction is a consequence of two elements, which were mentioned before: the magnetic profile of each vortex, determined by sharp boundary conditions, and the magnetic energy terms in the system that dominate over the condensation terms depending on the winding numbers. In agreement with the inversion of the direction of the current $J_0$ displayed in Fig~\ref{figure_current},
for $a' > 6.5$ the relative effective force reverts its direction.

Here, a critical case can be appreciated when  $\chi_-\to 1$, corresponding to coalescence of the vortices. This limit cannot be reached in our model due to the assumption that the vortices are widely separated.

\subsection{Equilibrium Position of the Vortices.}\label{VE}

From equation (\ref{43}) and Fig.~\ref{figure6}, the equilibrium angular position of the vortices is $\chi_- = 0$, corresponding to  $\gamma^*=-1$. Here, vortices have the largest separation between them in order to minimize the Helmholtz free energy of the system.

For $\gamma^*=-1$, the radial equilibrium positions of the vortices, $a_1^*$ and $a_2^*$, change with the size of the coaxial cylindrical boundaries. In order to illustrate the dependence between these variables, we keep fixed $R_0=4\lambda$ and we change $R$, for the cases $n_1=n_2=1$ and $n_1=1,n_2=-1$, respectively. 

\begin{figure}[h]
\includegraphics[width=1\columnwidth]{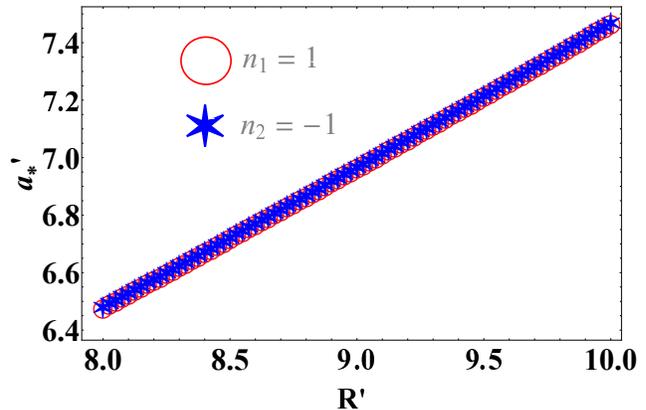}

\caption{Equilibrium positions of the first and the second vortex, as a function of the external radius of the annulus, for $n_1=n_2=1$ and $H_0=H_C$.}\label{figure7}
\end{figure}

Figure \ref{figure7} shows that the radial equilibrium positions of the vortices tend to move towards the external boundary as the size of the coaxial cylindrical region grows. This is a consequence of the mutual repulsion between vortices and the outer boundary of the sample. In other words, the exterior of the sample works as a giant pinning vortex, without superconducting electrons inside of it. 

\begin{figure}[h]
\includegraphics[width=1\columnwidth]{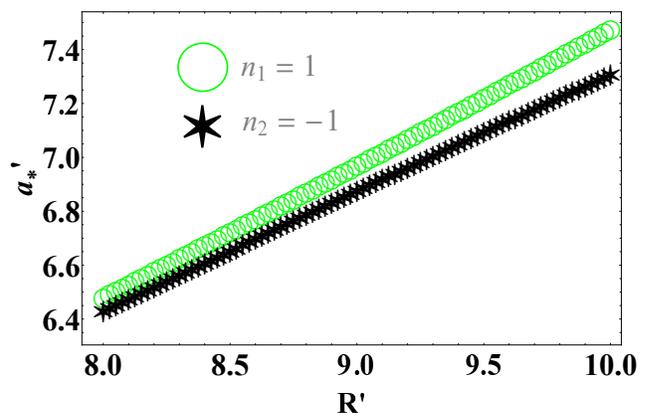}

\caption{Equilibrium positions of the first and the second vortex, as a function of the external radius of the annulus, for $n_1=1, n_2=-1$ and $H_0=H_C$.}\label{figure8}
\end{figure}

In Figs.~\ref{figure7} and \ref{figure8}, the radial equilibrium positions of the vortices present the same behavior, although in Fig.~\ref{figure8} the second vortex is closer to the center than the first one, due to the conservation of the fluxoid. This property can be checked using the classical analogue with the mutually inducting coils mentioned before. 

\section*{Conclusions}

In conclusion, our model predicts the repulsive interaction between single vortices in extreme Type II superconductivity, in agreement with the experiments and the theoretical developments until today. These results are obtained by solving for the magnetic profile everywhere, 
including the interior of each vortex. Our model preserves the convex shape of the general Ginzburg-Landau free energy, thus allowing for the search of an equilibrium configuration of the system as an absolute minima of the functional. We find that the angular equilibrium positions of the vortices are symmetrically related to cylindrical geometry of the sample, and the radial equilibrium positions are constrained by the fluxoid's conservation. In general, vortices maximize their distance when they come to the equilibrium,
in correspondence with an effective repulsive force. This last conclusion is also supported by a direct calculation of the thermodynamic surface free energy within our model.

\ack 
R.B. and D.G. thank the support of Iniciativa Cient\'ifica Milenio, N\'ucleo de F\'isica Matem\'atica, RC-12002 and FONDECYT 1160856. E.M. acknowledges FONDECYT 1190361.  

\appendix

\section{Magnetic Vector Potential and Magnetic Field in the Superconducting Domain}\label{B}

With the ansatz (\ref{17}), Ampere's Law for the superconducting domain can be written using equation (\ref{16}) and the rotational symmetry as 

\begin{eqnarray}
    \nabla^2A_0-\frac{A_0}{r^2}&=&-\frac{4\pi q^{*}\hbar\psi_{\infty}^2}{m^{*}cr}+\frac{4\pi(q^*)^2\psi_\infty^2A_0}{m^*c^2}.\label{B1}
\end{eqnarray}
In terms of the fluxoid, the coherence length and the penetration depth, Eq.~(\ref{B1}) can be written as: 

\begin{eqnarray}
    -\frac{\Phi_0}{2\pi\lambda^2r}&=&\frac{\partial^2A_0}{\partial r^2}+\frac{1}{r}\frac{\partial A_0}{\partial r}-A_0\left(\frac{1}{r^2}+\frac{1}{\lambda^2}\right).\label{B2}
\end{eqnarray}

The particular solution for (\ref{B2}) is given by 

\begin{equation}
A_0^p=\frac{\Phi_0}{2\pi r},\label{B3}
\end{equation}

\noindent
and with the change of variables $t=r\lambda^{-1}$, the homogeneous solution for (\ref{B2}) satisfies the Modified Bessel Equation  \cite{Bowman}: 

\begin{equation}
t^2\frac{\partial^2A_0}{\partial t^2}+t\frac{\partial A_0}{\partial t}-A_0\left(t^2+1\right)=0.\label{B4}
\end{equation}

Therefore, the magnetic vector potential inside the superconducting domain is given by 

\begin{equation}
\vec{A}_0=\left(c_1I_1\left(\frac{r}{\lambda}\right)+c_2K_1\left(\frac{r}{\lambda}\right)+\frac{\Phi_0}{2\pi r}\right)\hat{\theta}\label{B5}
\end{equation}

\noindent
and with the raising and lowering relations for Modified Bessel Functions \cite{Watson}, the magnetic field inside the superconducting domain is given by 

\begin{equation}
    \vec{B}_0=\frac{1}{\lambda}\left(c_1I_0\left(\frac{r}{\lambda}\right)-c_2K_0\left(\frac{r}{\lambda}\right)\right)\hat{k}.\label{B6}
\end{equation}

\section{Magnetic Vector Potential and Magnetic Field Inside the Vortex Domain}\label{C}

Inside each vortex domain $s_k \in \Omega_k$, we develop a scaling of the form $q_k=w_k\epsilon^{-\tau_k}$, for $k=1,2$ and $w_k = s_k/\lambda$. Here, $\tau$ is a scaling parameter that needs to be found. Then, Eq.(\ref{26}) takes the form 

\begin{eqnarray}
&-&\frac{\Phi_0n_k\epsilon^{\tau_k(2|n_k|+1)}q_k^{2|n_k|+1}}{2\pi \lambda\epsilon^{2|n_k|}}+\frac{A_{k,I}q_k^{2(|n_k|+1)}}{\epsilon^{2|n_k|}\epsilon^{-2\tau_k(|n_k|+1)}}\nonumber\\
&=&q_kA_{k,I}-A_{k,I}+q_k^2A_{k,I}''.\label{C1}
\end{eqnarray}

If $\tau_k=2|n_k|(2|n_k|+1)^{-1}$, after dropping negligible terms, (\ref{C1}) can be reduced to 

\begin{eqnarray}
-\left(\frac{\Phi_0n_k}{2\pi \lambda}\right)q_k^{2|n_k|+1} &=&q_k^2A_{k,I}''
+q_kA_{k,I}'-A_{k,I}.\label{C2}
\end{eqnarray}

The particular solution for (\ref{C2}) is given by

\begin{equation}
A_{k,I}^{P}=-\frac{\Phi_0n_kq_k^{2|n_k|+1}}{8\pi\lambda |n_k|(1+|n_k|)} \qquad k=1,2.\label{C3}
\end{equation}

Besides, the homogeneous solution for (\ref{C2}) is 

\begin{equation}
A_{k,I}^{H}=d_{k,I}q_k+\frac{e_{k,I}}{q_k}\qquad k=1,2.\label{C4}
\end{equation}

Therefore, using the fundamental relation $\vec{B}=\vec{\nabla}\times \vec{A}$, the magnetic vector potential and the magnetic field inside vortex $k$, for $k=1,2$, is given by 

   \begin{eqnarray}
\vec{A}_{k,I}&=&\left(\frac{d_{k,I}s_k}{\lambda\epsilon^{\tau_k}}+\frac{e_{k,I}\epsilon^{\tau_k}\lambda}{s_k}\right)\hat{\phi}_k\nonumber\\
&-&\left(\frac{\Phi_0n_ks_k^{2|n_k|+1}}{8\pi\lambda^2 |n_k|(1+|n_k|)\xi^{2|n_k|}}\right)\hat{\phi}_k,\label{C5}\\
\vec{B}_{k,I}&=&\left(\frac{2d_{k,I}}{\lambda\epsilon^{\tau_k}}-\frac{\Phi_0n_ks_k^{2|n_k|}}{4\pi\lambda^2|n_k|\xi^{2|n_k|}}\right)\hat{k}.\label{C6}
\end{eqnarray}

\section{Self Consistent Magnetic Flux}\label{D}

A vortex of radius $\xi<<\lambda$, located in $\vec{r}=\vec{a}$, is affected by the magnetic flux generated by an external magnetic potential of the form $\vec{A}=A(r)\hat{\theta}$. Then, the magnetic flux through the vortex, with internal coordinates $(s,\phi)$ and internal magnetic vector potential $\vec{A}_v=A_v(s)\hat{\phi}$, satisfies the condition: 

\begin{eqnarray}
\oint_{\partial\Omega_v}\vec{A}_{v}\cdot \vec{dl}&=&2\pi\xi A_{v}(s=\xi)\label{D1}\\
&=&\oint_{\partial\Omega_v}\vec{A}\cdot \vec{dl}\nonumber\\
&=&\int_{0}^{2\pi}\xi A\left(|\vec{a}+\vec{\xi}|\right)(\hat{\theta}\cdot \hat{\phi})d\phi\nonumber\\
&\simeq& \int_{0}^{2\pi}\xi A\left(a\right)(\sin\theta\sin\phi+\cos\theta\cos\phi)\nonumber\\
&=& \int_{0}^{2\pi}\left(\frac{\xi A\left(a\right)(\xi+a\cos\phi-\alpha)}{\sqrt{\xi^2+2a\xi\cos\phi-\alpha+a^2}}\right)d\phi.\nonumber
\end{eqnarray}

In the limit $\xi<<a$, the last equation can be written as 

\begin{eqnarray}
\oint_{\partial\Omega_v}\vec{A}_v\cdot \vec{dl}&\simeq&\int_{0}^{2\pi}\xi A\left(a\right)\left(\frac{\xi}{a}+\cos\phi-\alpha\right)d\phi\label{D2}\\
&-&\int_{0}^{2\pi}\left(\frac{\xi^3A(a)\cos\phi-\alpha}{a^2}\right)d\phi\nonumber\\
&-&\int_{0}^{2\pi}\left(\frac{\xi^2A(a)\cos^2\phi-\alpha}{a}\right)d\phi\nonumber\\
&=&\frac{\xi^2\pi}{a}A(a).\nonumber
\end{eqnarray}

Therefore, we conclude that

\begin{equation}
    \frac{2A_v(s=\xi)}{\xi}\simeq \frac{A(r=a)}{a}.\label{D3}
\end{equation}

\section{Expressions for the unknown constants of the problem}\label{Cnt}

Defining the following function for the distance between the vortex's centers: 

\begin{equation}
    a_{1,2}:=|\vec{a}_1-\vec{a}_2|,\label{Cn1}
\end{equation}

\noindent 
and the auxiliary functions: 

\begin{eqnarray}
    \mathcal{A}_1(\vec{a}_1,\vec{a}_2)&:=&\frac{\Phi_0n_1}{4\pi\lambda |n_1|}\left(\frac{1}{1+|n_1|}-1\right)\nonumber\\
    &+&\frac{\lambda\Phi_0n_2}{2\pi a_{1,2}^2}+\frac{\lambda A_0(a_1)}{a_1},\label{Cn2}\\
\mathcal{A}_2(\vec{a}_1,\vec{a}_2)&:=&\frac{\Phi_0n_2}{4\pi\lambda |n_2|}\left(\frac{1}{1+|n_2|}-1\right)\nonumber\\
&+&\frac{\lambda\Phi_0n_1}{2\pi a_{1,2}^2}+\frac{\lambda A_0(a_2)}{a_2},\label{Cn3}\\
\mathfrak{B}_1(\vec{a}_1,\vec{a}_2)&:=&\lambda\left(I_0(\epsilon)B_0(a_2)-\frac{\lambda I_1\left(\frac{a_{1,2}}{\lambda}\right)B_0(a_1)}{a_{1,2}}\right)\nonumber\\
&+&I_0\left(\frac{a_{1,2}}{\lambda}\right)\mathcal{A}_1-I_0(\epsilon)\mathcal{A}_2\label{Cn4},\\
\mathfrak{B}_2(\vec{a}_1,\vec{a}_2)&:=&\lambda\left(I_0(\epsilon)B_0(a_1)-\frac{\lambda I_1\left(\frac{a_{1,2}}{\lambda}\right)B_0(a_2)}{a_{1,2}}\right)\nonumber\\
&+&I_0\left(\frac{a_{1,2}}{\lambda}\right)\mathcal{A}_2-I_0(\epsilon)\mathcal{A}_1.\label{Cn5}\\
\mathsf{e}_1(\vec{a}_1,\vec{a}_2)&:=&I_0(\epsilon)\left(K_0\left(\frac{a_{1,2}}{\lambda}\right)+\frac{\lambda}{a_{1,2}}K_1\left(\frac{a_{1,2}}{\lambda}\right)\right)\nonumber\\
&-&K_0(\epsilon)\left(I_0\left(\frac{a_{1,2}}{\lambda}\right)-\frac{\lambda}{a_{1,2}}I_1\left(\frac{a_{1,2}}{\lambda}\right)\right),  \label{Cn6}\\
\mathsf{e}_2(\vec{a}_1,\vec{a}_2)&:=&\frac{\lambda}{a_{1,2}}I_0\left(\frac{a_{1,2}}{\lambda}\right)K_1\left(\frac{a_{1,2}}{\lambda}\right)\nonumber\\
&+&\frac{\lambda}{a_{1,2}}I_1\left(\frac{a_{1,2}}{\lambda}\right)K_0\left(\frac{a_{1,2}}{\lambda}\right),\label{Cn7}\\
\mathtt{E}(\vec{a}_1,\vec{a}_2)&:=&\frac{\mathsf{e}_1(\vec{a}_1,\vec{a}_2)}{\left(\mathsf{e}_1(\vec{a}_1,\vec{a}_2)\right)^2-\left(\mathsf{e}_2(\vec{a}_1,\vec{a}_2)\right)^2},\label{Cn8}\\
\mathtt{F}(\vec{a}_1,\vec{a}_2)&=&\frac{\mathsf{e}_2(\vec{a}_1,\vec{a}_2)}{\left(\mathsf{e}_1(\vec{a}_1,\vec{a}_2)\right)^2-\left(\mathsf{e}_2(\vec{a}_1,\vec{a}_2)\right)^2}\label{Cn9}
\end{eqnarray}

\noindent 
the constants related to the boundary conditions can be written as: 

\begin{eqnarray}
    e_{1,E}&=&\mathfrak{B}_1\mathtt{E}+\mathfrak{B}_2\mathtt{F},\label{Cn10}\\
    e_{2,E}&=&\frac{\mathfrak{B}_2+\mathsf{e}_2e_{1,E}}{\mathsf{e}_1},\label{Cn11}\\
    d_{1,E}&=&\frac{\mathcal{A}_2-\lambda B_0(a_2)}{I_0\left(\frac{a_{1,2}}{\lambda}\right)-\frac{\lambda}{a_{1,2}}I_1\left(\frac{a_{1,2}}{\lambda}\right)}\nonumber\\
    &+&\left(\frac{K_0\left(\frac{a_{1,2}}{\lambda}\right)+\frac{\lambda}{a_{1,2}}K_1\left(\frac{a_{1,2}}{\lambda}\right)}{I_0\left(\frac{a_{1,2}}{\lambda}\right)-\frac{\lambda}{a_{1,2}}I_1\left(\frac{a_{1,2}}{\lambda}\right)}\right)e_{1,E}\label{Cn12},\\
    d_{2,E}&=&\frac{\mathcal{A}_1-\lambda B_0(a_1)}{I_0\left(\frac{a_{1,2}}{\lambda}\right)-\frac{\lambda}{a_{1,2}}I_1\left(\frac{a_{1,2}}{\lambda}\right)}\nonumber\\
    &+&\left(\frac{K_0\left(\frac{a_{1,2}}{\lambda}\right)+\frac{\lambda}{a_{1,2}}K_1\left(\frac{a_{1,2}}{\lambda}\right)}{I_0\left(\frac{a_{1,2}}{\lambda}\right)-\frac{\lambda}{a_{1,2}}I_1\left(\frac{a_{1,2}}{\lambda}\right)}\right)e_{2,E},\label{Cn13}\\
    d_{1,I}&=&\frac{\epsilon^{\tau_1}}{2}\left[\frac{\Phi_0n_1}{4\pi\lambda|n_1|}+d_{1,E}I_0(\epsilon)-e_{1,E}K_0(\epsilon)\right],\label{Cn14}\\
    d_{2,I}&=&\frac{\epsilon^{\tau_2}}{2}\left[\frac{\Phi_0n_2}{4\pi\lambda|n_2|}+d_{2,E}I_0(\epsilon)-e_{2,E}K_0(\epsilon)\right].\label{Cn15}
\end{eqnarray}
\section{Computation of the Helmholtz Free Energy Profile}\label{E}

In fact, the Helmholtz free energy functional only contains terms that depend on $\vec{a}_1$ and $\vec{a}_2$, and are related to the magnetic vector potentials and the magnetic fields of each vortex. Therefore, for $k=1,2$, the first relevant term is: 

\noindent 
\begin{eqnarray}
\delta f_{1,k}&=&\int_{\Omega_k}\frac{|\vec{B}_{k,I}|^2}{8\pi}d^2x.\label{E1}
\end{eqnarray}
Substituting the internal magnetic field profile of the vortex, we have
\begin{eqnarray}
4\delta f_{1,k}&=&\int_0^{\xi}s_k\left[\frac{2d_{k,I}}{\lambda\epsilon^{\tau_k}}-\frac{\Phi_0n_ks_k^{2|n_k|}}{4\pi\lambda^2|n_k|\xi^{2|n_k|}}\right]^2ds_k\label{E2}\\
&=&\int_0^{\xi}\frac{4d_{k,I}^2s_kds_k}{\lambda^2\epsilon^{2\tau_k}}\nonumber\\
&-&\int_0^{\xi}\frac{\Phi_0n_ks_k^{2|n_k|+1}d_{k,I}ds_k}{\pi\lambda^3|n_k|\epsilon^{\tau_k}\xi^{2|n_k|}}\nonumber\\
&+&\int_0^{\xi}\frac{\Phi_0^2s_k^{4|n_k|+1}ds_k}{16\pi^2\lambda^4\xi^{4|n_k|}}\nonumber\\
&=&\left.\frac{4d_{k,I}^2s_k^2}{2\lambda^2\epsilon^{2\tau_k}}-\frac{\Phi_0n_ks_k^{2|n_k|+2}d_{k,I}}{2\pi\lambda^3|n_k|(1+|n_k|)\epsilon^{\tau_k}\xi^{2|n_k|}}\right|_0^{\xi}\nonumber\\
&+&\left.\frac{\Phi_0^2s_k^{4|n_k|+2}}{32\pi^2\lambda^4\xi^{4|n_k|}(1+2|n_k|)}\right|_0^{\xi}\nonumber.
\end{eqnarray}

Thus,

\begin{eqnarray}
\delta f_{1,k}&=&\frac{\epsilon^2d_{k,I}^2}{2\epsilon^{2\tau_k}}-\frac{\Phi_0n_k\epsilon^2d_{k,I}}{8\pi\lambda\epsilon^{\tau_k}|n_k|(1+|n_k|)}\nonumber\\
&+&\frac{\Phi_0^2\epsilon^2}{128\pi^2\lambda^2(1+2|n_k|)}.\label{E4}
\end{eqnarray}

For the next relevant term of the energy: 

\begin{eqnarray}
\fl\frac{2m^{*}c^2\delta f_{2,k}}{(q^{*})^2}&=&\left(\int_{\Omega_k}|\psi_{k,I}|^2|\vec{A}_{k,I}|^2d^2x\right)\nonumber\\
&=&\left(\int_{0}^{2\pi}\int_0^{\xi}|\psi_{k,I}|^2|\vec{A}_{k,I}|^2s_kds_kd\phi_k\right)\label{E5}
\end{eqnarray}

\noindent
which leads us to 

\begin{eqnarray}
\delta f_{2,k}&=&\int_0^{\xi}\frac{d_{k,I}^2s_k^{2|n_k|+3}ds_k}{4\lambda^4\epsilon^{2\tau_k}\xi^{2|n_k|}}\nonumber\\
&-&\int_0^{\xi}\frac{\Phi_0n_kd_{k,I}s_k^{4|n_k|+3}ds_k}{16\pi\lambda^5\epsilon^{\tau_k}|n_k|(1+|n_k|)\xi^{4|n_k|}}\nonumber\\
&+&\int_0^{\xi}\frac{\Phi_0^2s_k^{6|n_k|+3}ds_k}{256\pi^2\lambda^6(1+|n_k|)^2\xi^{6|n_k|}}\label{E6}\\
&=&\frac{d_{k,I}^2\xi^{4}}{8\lambda^4\epsilon^{2\tau_k}(2+|n_k|)}-\frac{\Phi_0n_kd_{k,I}\xi^{4}}{64\pi\lambda^5\epsilon^{\tau_k}|n_k|(1+|n_k|)^2}\nonumber\\
&+&\frac{\Phi_0^2\xi^{4}(2+3|n_k|)^{-1}}{512\pi^2\lambda^6(1+|n_k|)^2}\label{E7}.
\end{eqnarray}

Then,

\begin{eqnarray}
\delta f_{2,k}&=&\frac{\epsilon^4(d_{k,I})^2}{8(2+|n_k|)\epsilon^{2\tau_k}}-\frac{\Phi_0n_k\epsilon^4d_{k,I}}{64\pi\lambda|n_k|(1+|n_k|)^2\epsilon^{\tau_k}}\nonumber\\
&+&\frac{\Phi_0^2\epsilon^4}{512\pi^2\lambda^2(2+3|n_k|)(1+|n_k|)^2}.\label{E8} 
\end{eqnarray}

And the last relevant term is related to 

\begin{eqnarray}
-\frac{2m^{*}ic\delta f_{3,k}}{q^{*}\hbar}&=&\int_{\Omega_k} \vec{A}_{k,I}\cdot\psi_{k,I}^{*}\vec{\nabla}_k\psi_{k,I}d^2x\nonumber\\
&-&\int_{\Omega_k} \vec{A}_{k,I}\cdot\psi_{k,I}\vec{\nabla}_k\psi_{k,I}^{*}d^2x\label{E9}\\
\end{eqnarray}
Substituting for the order parameter solution inside the vortex, we obtain
\begin{eqnarray}
-\frac{m^{*}c\xi^{2|n_k|}\delta f_{3,k}}{2\pi q^{*}\hbar \psi_{\infty}^2n_k}&=&\int_0^{\xi}\frac{d_{k,I}s^{2|n_k|+1}ds_k}{\lambda\epsilon^{\tau_k}}\nonumber\\
&-&\int_0^{\xi}\frac{\Phi_0n_ks_k^{4|n_k|+2}ds_k}{8\pi\lambda^2|n_k|(1+|n_k|)\xi^{2|n_k|}}\label{E10}\\
&=&\frac{d_{k,I}\xi^{2|n_k|+2}}{2\lambda(1+|n_k|)\epsilon^{\tau_k}}\nonumber\\&-&\frac{\Phi_0n_k\xi^{2|n_k|+2}}{16\pi\lambda^2|n_k|(1+|n_k|)(1+2|n_k|)}\nonumber.
\end{eqnarray}

Thus,
\begin{eqnarray}
\delta f_{3,k}&=&\frac{\Phi_0^2n_k^2\epsilon^2}{64\pi^2\lambda^2|n_k|(1+|n_k|)(1+2|n_k|)}\nonumber\\
&-&\frac{\Phi_0\epsilon^2n_kd_{k,I}}{8\pi\lambda(1+|n_k|)\epsilon^{\tau_k}}.\label{E11}
\end{eqnarray}

\section{Surface Energy}\label{F}

In the same spirit of \cite{Fetter}, we compute and approximation to the Gibbs free energy of the interfaces at $H_0=H_C$: 

\begin{equation}
    \sigma_{ns}=\frac{H_C^2}{8\pi}\int_{\Omega}\left(\left\{1-\frac{B}{H_C}\right\}^2-\frac{|\psi|^4}{\psi_{\infty}^4}\right)d^2x,\label{F1}
\end{equation}
Considering the geometrical structure of the domain $\Omega$, we have
\begin{eqnarray}
&&\frac{8\pi\sigma_{ns}}{H_C^2}=\int_{\Omega\setminus(\Omega_1\cup\Omega_2)}\left(\left\{1-\frac{B_0(r)}{H_C}\right\}^2-\frac{|\psi_0(r)|^4}{\psi_{\infty}^4}\right)d^2x\nonumber\\
&+&\sum_{k=1}^2\int_{\Omega_k}\left(\left\{1-\frac{B_{k,I}(s_k)}{H_C}\right\}^2-\frac{|\psi_{k,I}(s_k)|^4}{\psi_{\infty}^4}\right)d^2x.\label{F2}
\end{eqnarray}

In the limit $\kappa>>1$, if we define the following integrals as : 

\begin{eqnarray}
J_1&=&\frac{R^2-R_0^2}{2}-\frac{2R}{H_C}\left(c_1I_0\left(\frac{R}{\lambda}\right)+c_2K_0\left(\frac{R}{\lambda}\right)\right)\nonumber\\
&+&\frac{2R_0}{H_C}\left(c_1I_0\left(\frac{R_0}{\lambda}\right)+c_2K_0\left(\frac{R_0}{\lambda}\right)\right)\nonumber\\
&+&\frac{c_1^2R^2}{2H_C^2\lambda^2}\left(I_0^2\left(\frac{R}{\lambda}\right)-I_1^2\left(\frac{R}{\lambda}\right)\right)\nonumber\\
&-&\frac{c_1^2R_0^2}{2H_C^2\lambda^2}\left(I_0^2\left(\frac{R_0}{\lambda}\right)-I_1^2\left(\frac{R_0}{\lambda}\right)\right)\nonumber\\
&+&\frac{c_2^2R^2}{2H_C^2\lambda^2}\left(K_0^2\left(\frac{R}{\lambda}\right)-K_1^2\left(\frac{R}{\lambda}\right)\right)\nonumber\\
&-&\frac{c_2^2R_0^2}{2H_C^2\lambda^2}\left(K_0^2\left(\frac{R_0}{\lambda}\right)-K_1^2\left(\frac{R_0}{\lambda}\right)\right)\nonumber\\
&-&\frac{c_1c_2}{H_C^2}\int_{\frac{R_0}{\lambda}}^{\frac{R}{\lambda}}uI_0(u)K_0(u)du,\label{F3}
\end{eqnarray}

\begin{eqnarray}
J_2&\simeq&\xi^2\left(1-\frac{1}{H_C\lambda}\sum_{k=1}^2\left\{c_1I_0\left(\frac{a_k}{\lambda}\right)-c_2K_0\left(\frac{a_k}{\lambda}\right)\right\}\right)\nonumber\\
&+&\frac{\xi^2}{2\lambda^2H_C^2}\sum_{k=1}^2\left(\left\{c_1I_0\left(\frac{a_k}{\lambda}\right)-c_2K_0\left(\frac{a_k}{\lambda}\right)\right\}^2\right),\label{F4}
\end{eqnarray}

\begin{eqnarray}
   J_3&=&\sum_{k=1}^2\frac{\xi^2}{2}\left(1-\frac{4d_{k,I}}{\lambda H_C\epsilon^{\tau_k}}\right)+\frac{\Phi_0n_k\xi^2}{4\pi\lambda^2H_C|n_k|(|n_k|+1)}\nonumber\\
   &+&\frac{2d_{k,I}^2\xi^2}{H_C^2\lambda^2\epsilon^{2\tau_k}}-\frac{\Phi_0n_kd_{k,I}\xi^2}{2\pi\lambda^3H_C^2\epsilon^{\tau_k}|n_k|(|n_k|+1)}\nonumber\\
   &+&\frac{\Phi_0^2\xi^2}{32\pi^2\lambda^4H_C^2(2|n_k|+1)}.\label{F5}
\end{eqnarray}
In terms of the expressions above, we define the parameters
\begin{eqnarray}
    \frac{\gamma_1}{\pi}&=&R_0^2-R^2-2\lambda(R+R_0)+2(J_1-J_2+J_3),\label{F6}\\
    \frac{\gamma_2}{\pi}&=&-4\xi(R+R_0)+2\xi^2\sum_{k=1}^2\left(\frac{1}{2|n_k|+1}\right).\label{F7}
\end{eqnarray}

In terms of the definitions above, the surface energy is: 

\begin{equation}
    \sigma_{ns}\simeq\left(\frac{H_C^2}{8\pi}\right)\left(\gamma_1-\gamma_2\right),\label{F8}
\end{equation}

\noindent
Equation (\ref{F8}) can be approximated to a value that does not depend of each vortex's position, as presented for instance in \cite{Tinkham}: 

\begin{eqnarray}
    \sigma_{ns}&\simeq&\frac{H_C^2\xi^2(1-\kappa)}{2}\left(2+\frac{(R+R_0)}{\xi}\right)\label{F9}\\
    &<&0.\nonumber
\end{eqnarray}


%
%
%
%
%

\end{document}